\documentclass[pra,twocolumn,showpacs,letterpaper,showpacs,superscriptaddress]{revtex4}
\usepackage{hyperref}
\usepackage{graphicx}
\usepackage{amsmath}

\begin{document}
\newcommand{\eq}[1]{Eq.~(\ref{#1})}
\newcommand{\bE}{{\bf E}}
\newcommand{\bD}{{\bf D}}
\newcommand{\bH}{{\bf H}}
\newcommand{\bB}{{\bf B}}
\newcommand{\bS}{{\bf S}}
\newcommand{\br}{{\bf r}}

\newcommand{\lc}{$\ell$-cavity}
\newcommand{\Lc}{$L$-cavity}
\newcommand{\llc}{$c$-cavity}

\title{Vacuum field energy and spontaneous emission in anomalously dispersive cavities}
\author{Douglas H. Bradshaw}
\email{bradshaw@lanl.gov}
\author{Michael D. Di Rosa}
\affiliation{Los Alamos National Laboratory, Los Alamos, NM 87545 USA}
\date{\today}

\begin{abstract}
Anomalously dispersive cavities, particularly white light cavities, may have larger bandwidth to finesse ratios than their normally dispersive counterparts.  Partly for this reason, their use has been proposed for use in LIGO-like gravity wave detectors and in ring-laser gyroscopes. In this paper we analyze the quantum noise associated with anomalously dispersive cavity modes.
The vacuum field energy associated with a particular cavity mode is proportional to the cavity-averaged group velocity of that mode.
For anomalously dispersive cavities with group index values between 1 and 0, this means that the total vacuum field energy associated with a particular cavity mode must exceed $\hbar \omega/2$.
For white light cavities in particular, the group index approaches zero and the vacuum field energy of a particular spatial mode may be significantly enhanced.
We predict enhanced spontaneous emission rates into anomalously dispersive cavity modes and broadened laser linewidths when the linewidth of intracavity emitters is broader than the cavity linewidth.
\end{abstract}
\maketitle
\section{Introduction}\label{se:introduction}

When the finesse of an optical cavity is improved, buildup \footnote{By buildup we refer to the resonant enhancement of intracavity intensity as normalized by the incident intensity of a mode.  For high-finesse cavities where the round trip electric field transformation is $E\rightarrow E e^{-i\phi}(1-\delta)$, the buildup is $4/\delta^2$.} increases but resonance bandwidth decreases.  
Resonance bandwidth may be increased without degrading buildup by decreasing cavity length.
For some applications, such as interferometry-based gravity-wave detection and Sagnac interferometry, decreasing cavity length degrades overall system performance.  
For other applications, such as fast single-photon generation \cite{Englund-Fattal-Vuckovic-2005,Moerk-2010}, decreasing cavity length may be profitable, but is not possible past a minimum length.

When actual decreases in length are either unprofitable or impossible, it is natural to consider the use of anomalous intra-cavity dispersion.  
For a given cavity geometry, buildup scales with finesse ($F$), while resonance bandwidth scales inversely with quality ($Q$).  
Quality and finesse are related by $Q=F\times(\nu_o L_g/c)$, where $\nu_o$ is the resonant frequency, $c$ is the speed of light, and $L_g$ is the round-trip group optical path difference.  $L_g$ is defined in terms of the round trip phase $\phi$ and the angular frequency $\omega$ by  $L_g \equiv c d\phi/d\omega$ \cite{Candler-1946}.
In the presence of dispersion, $L_g =\ell n_g$, where $\ell$ is the cavity round trip length and $n_g$ is an effective cavity group index.  The response of the ratio $F/Q$ to changes in $n_g$ is identical to its response to changes in $\ell$.
However, while $\ell$, has a minimum resonant value ($\lambda/2$ for symmetric cavities), $n_g$ may approach zero for a particular range of frequencies \cite{Wicht-Muller-Rinkleff-Danzmann-2000,Pati-Salit-Salit-Shahriar-2007}.
Cavities where $n_g$ approaches zero at a cavity resonance have been referred to as ``white-light cavities,'' in reference to their dispersion-broadened resonances \cite{Wicht-Danzmann-Rinkleff-1997}.  

The white-light cavity literature is centered around two potential applications, gravity wave sensing \cite{Wicht-Danzmann-Rinkleff-1997, Wise-Whiting-2004,Karapetyan-2004} and rotation sensing \cite{Shahriar-Pati-Tripathy-Gopal-Messall-Salit-2007}.  LIGO-like gravity wave detectors take the form of large Michelson interferometers. 
In signal recycling, mirrors may be placed near the entrance to each Michelson arm, forming a pair of Gire-Tournois cavities.
In power recycling, a mirror is placed in advance of the central beam splitter, making the entire interferometer into an optical cavity.
In either case, the increase in signal due to enhanced buildup is gained only at the expense of bandwidth, and
white-light cavities have been considered as a way to combine large bandwidth with large buildup.

Rotation in a ring-resonator gyroscope is equivalent to cavity elongation/truncation when resonator and mode are co-/counter-rotating \cite{Shahriar-Pati-Tripathy-Gopal-Messall-Salit-2007}.  
The frequency response of a cavity mode to a change in length may be enhanced via anomalous dispersion;
as the round-trip length $\ell$ changes, the resonant frequency $\omega_o$ associated with a given mode changes according to $d\omega_o/d\ell=-n\omega_o/(\ell n_g)$, which grows large as $n_g$ approaches $0$.
Shahriar \emph{et al.} found that the increase of frequency sensitivity in a passive anomalously dispersive ring cavity would be counterbalanced by the concurrent increase in resonance bandwidth, leaving no net gain in rotation resolution
\cite{Shahriar-Pati-Tripathy-Gopal-Messall-Salit-2007}. 
However, they argued that this cancellation could be avoided by using an active interferometer in the form of a white-light ring-laser gyroscope.

The effectiveness of white-light cavities for any sensing application is partially determined by white-light cavity noise.
Some implementation-dependent sources of noise have been quantified by Wicht \emph{et al.} \cite{Wicht-Danzmann-Rinkleff-1997} for a double-lambda system and by Sun \emph{et al.} \cite{Sun-Shahriar-Zubairy-2009} for a double gain system.  
A comparison of several implementations can be found in a later work by Wicht \emph{et al.} \cite{Wicht-Rinkleff-Danzmann-2002}.
However, the white-light cavity literature does not include, to our knowledge, any work on the noise due to vacuum energy intrinsic to a white-light cavity.

In this paper, we show by simple and largely classical arguments that the vacuum field energy associated with the ground state of an anomalously dispersive cavity mode diverges as that mode approaches the ideal ($n_g\rightarrow 0, F\rightarrow \infty$) associated with a white-light cavity.  
This divergence is intrinsic to the definition of a white-light cavity and is independent of any specific implementation. It has the consequence that spontaneous emission of an excited particle into the white-light cavity mode may increase substantially.  
This, in turn, has two consequences.  
First, the width of a white-light laser line must be broadened by anomalous dispersion.  
This broadening, if applied to the configuration proposed by Shahriar \emph{et al.} \cite{Shahriar-Pati-Tripathy-Gopal-Messall-Salit-2007} would 
cancel the increased frequency sensitivity of a white-light ring-laser gyroscope.
Second, it suggests that anomalous dispersion may be useful in enhancing the quantum yield of single photon emitters where radiative decay into the mode of interest must compete with nonradiative decay or with radiation into other spatial modes.

In the following section we review the result of applying the standard classical expression for the narrow-bandwidth energy density in a lossless, dispersive medium to the quantization of a closed cavity.  We introduce a distinction between two energies, which we call the ``field energy'' and the ``total energy,'' and which are related by the ratio between cavity-averaged phase and group velocities.  The total cavity energy is quantized as a simple harmonic oscillator, but only the field energy interacts with dipoles inside the cavity.
In Section III, we explore the vacuum field noise in an anomalously dispersive cavity using an alternative coupled-cavity approach.  This approach allows us to find an expression for the vacuum fields that is independent of any specific expression for the electromagnetic energy density in a dispersive medium.  It also provides a simple view of the effect of dispersion on the power spectrum of the vacuum field energy.  We find that this approach also recovers the results of Section II when the assumptions of that section (no loss and negligible group velocity dispersion) are applied.  We then relax the assumptions of Section II and use a numerical model of a physical medium and show that as the round trip loss approaches zero, the quantum field noise of a white light cavity resonance diverges.  We briefly address the relationship between this increased noise and the quantum limited laser linewidth of anomalously dispersive cavities.

\section{Spontaneous emission in a lossless, dispersive medium}\label{se:closed}
Electromagnetic field quantization in an evacuated, closed and lossless cavity proceeds as follows:  
\begin{enumerate}
\item A complete set of orthonormal electromagnetic modes is identified.  
\item The volume-integrated energy of each of these modes is represented.
\item Each mode is quantized as though it were a simple harmonic oscillator.  \footnote{Note that this is the only step that introduces a non-classical element.  It is the two classical steps that define the relationship between mode excitations and electromagnetic fields.}
\end{enumerate}

When a dispersive medium is introduced into a cavity, both steps 1 and 2 become problematic.   Step 1 becomes problematic because a dispersive medium must have loss (or gain), which couples the electromagnetic fields of a dispersive cavity to external degrees of freedom.  True electromagnetic modes can then only be found by explicitly including these external degrees of freedom. 

Step 2 becomes problematic because dispersion also complicates the concept of electromagnetic field energy density.  Whether or not energy that has been stored by a dispersive medium returns to field form depends on future interactions between medium and field \footnote{a concrete application of this concept may be found in \cite{Glasgow-Meilstrup-Peatross-Ware-2007}}.  In other words, the electromagnetic energy density of a dispersive medium is nonlocal in time.

These difficulties with steps 1 and 2 can both be avoided if, rather than seek a complete set of orthonormal electromagnetic modes we focus on a single cavity mode.  Although a causal dispersive medium must have loss (or gain), causality does not prevent this loss from being small or even vanishing for particular frequencies.  Thus, although a causal dispersive cavity must be open (i.e. coupled to external degrees of freedom) in general, it may be effectively closed at particular resonances.  In addition, a perfect, closed cavity mode has a vanishing spectral width.  Although the electromagnetic energy density in a dispersive medium is, in general, an ill-defined quantity, it is well defined in the absence of loss for spectrally narrow excitations.

\subsection{Energy density of a quasimonochromatic planar wave in a lossless, dispersive medium}
The effect of dispersion on the electromagnetic energy density of a quasimonochromatic excitation in a lossless linear medium is particularly simple for planar waves.
When only radiative energy transportation is non-negligible, Poynting's theorem takes the form 
\begin{equation}
\nabla \cdot \bS=-\partial u/\partial t, 
\end{equation} 
where $\bS$ is the Poynting vector and $u$ is the electromagnetic energy density.
If absorption, scattering, and group velocity dispersion are all negligible, a direct calculation of the divergence of a spectrally narrow planar wave propagating through a dispersive medium yields
\begin{equation}\label{Eq:u-from-S}
 u(z,t)=\frac{\langle S(z,t)\rangle}{v_g},
\end{equation}
where $v_g$ is the group velocity of the planar wave, and the angle brackets denote a cycle average.

If dispersion were neglected, $\langle S\rangle/v_g$ in \eq{Eq:u-from-S} would become $\langle S\rangle /v_p$, where $v_p$ is the phase velocity.  In other words, if $u_f$ is an expression for the nondispersive energy density, then $u$ can be found from $u_f$ via
\begin{equation}\label{Eq:u-from-uf}
 u=\frac{n_g}{n} u_f,
\end{equation} 
where $n_g$, and $n$ are the refractive index and the group index.  The standard form for the cycle averaged electromagnetic energy density in a lossless, isotropic, linear, dispersionless medium is \cite{Landau-Pitaevski-Lifshitz, Jackson-1998}
\begin{equation}\label{Eq:uf}
 u_f=\frac{\epsilon \langle E^2\rangle }{2}+\frac{\mu \langle H^2\rangle}{2}.
\end{equation}
Substituting \eq{Eq:uf} into \eq{Eq:u-from-uf} gives the cycle averaged electromagnetic energy density in a lossless, isotropic, linear, dispersive medium,
\begin{equation}\label{Eq:u-planar}
 u=\frac{n_g}{n}\left(\frac{\epsilon \langle E^2\rangle }{2}+\frac{\mu \langle H^2\rangle}{2}\right).
\end{equation}

For spectrally narrow excitations in lossless media, dispersion has a simple relationship to energy storage by the dispersive medium.  If we call $u_f$ the field energy density and $u$ the total energy density then we can also define a third quantity, $u_s$ as the stored energy density, or the difference between the total energy density and the field energy density.  In normally dispersive media, $u>u_f$, and $u_s$ is positive because electromagnetic energy is dispersively stored in the medium.  In anomalously dispersive cavities, where $0\leq n_g <n$, $u<u_f$ and $u_s$ is negative.  Just as $u_s>0$ in a dispersive medium signifies temporary absorption, $u_s<0$ in an anomalously dispersive medium signifies temporary emission.  In either case, energy exchange is governed by the interaction between the instantaneous spectrum of the planar wave and the response function complex susceptibility of the dispersive medium \cite{Peatross-Ware-Glasgow-2001}.  In the limit of a white light cavity, $u_s=-u_f$ and $u=0$, meaning that all of the field energy has been donated by the medium.

A strength of \eq{Eq:u-from-S} and subsequent expressions is that they show the effect of dispersion on the energy density explicitly and simply.  However, these expressions were derived for planar waves and do not apply when there is interference between waves propagating in different directions.  This rules out their application to many optical cavities.

\subsection{Energy of a quasimonochromatic mode in a lossless, dispersive medium}
A more general but less intuitive form for the energy density may be found by using vector identities and Maxwell's equations to make the transformation
\begin{equation}
\nabla \cdot \bS=\bE\cdot \frac{\partial\bD}{\partial t}+\bH\cdot\frac{\partial \bB}{\partial t},
\end{equation}
and then using the narrow-band nature of the excitation and linearity to obtain \cite{Landau-Pitaevski-Lifshitz, Jackson-1998}
\begin{equation}\label{Eq:u-standard}
u = Re\left[\frac{d (\omega \epsilon)}{d\omega}\right]\frac{\langle E^2\rangle}{2}+Re\left[\frac{d (\omega \mu)}{d\omega}\right]\frac{\langle H^2 \rangle}{2},
\end{equation} 
where the angle brackets denote a cycle average.

\eq{Eq:u-standard} shows that the simple relationship between dispersion and energy density given for planar waves in \eq{Eq:u-from-uf} does not generalize to non-planar waves.  For example, consider the energy density of counter-propagating waves of equal amplitude and frequency and identical polarization in a dielectric. If we assume that $\epsilon$ and $\mu$ and their first frequency derivatives are real, then we can use the identities $n=\sqrt{\epsilon\mu/(\epsilon_o\mu_o)}$ and $n_g=n+\omega dn/d\omega$ to obtain
\begin{equation}\label{Eq:ng-over-n}
 \frac{n_g}{n}=1+\frac{1}{2}\frac{d \ln \epsilon}{d\ln \omega}+\frac{1}{2}\frac{d \ln \mu}{d \ln \omega}.
\end{equation}
Since the medium is a lossless dielectric, $\omega d\epsilon/d\omega=\eta\epsilon$, and $\omega d\mu/d\omega=0$, where $\eta$ is a real number.  Applying these values to \eq{Eq:ng-over-n} gives $n_g/n=1+\eta/2$. 
In the standing wave that results from the counter-propagating waves, electric field nodes correspond with magnetic field antinodes.  At electric field nodes, $E^2=0$, and $u/u_f=1$.  At electric field anti-nodes, $H^2=0$, and $u/u_f=1+\eta$.  Thus, \eq{Eq:uf} does not generalize to non-planar waves.

However, an analogue to \eq{Eq:uf} does apply to cavity modes.  
Assuming losslessness and vanishing imaginary derivatives for $\epsilon$ and $\mu$, we rewrite
\eq{Eq:u-standard} as 
\begin{equation}\label{Eq:u-lossless}
u = \left(1+\frac{d \ln \epsilon}{d\ln \omega}\right)\frac{\epsilon \langle E^2\rangle}{2}+\left(1+\frac{d \ln \mu}{d\ln \omega}\right)\frac{\mu \langle H^2 \rangle}{2}.
\end{equation} 

We now note that the electric and magnetic contributions to the total modal energy of a nondispersive cavity mode are equal.  That is, for the field associated with a mode that satisfies closed boundary conditions and the Helmholtz equation,
\begin{equation}\label{Eq:yin=yang}
 \int \frac{\epsilon \langle E^2\rangle}{2}d^3 r=\int \frac{\mu \langle H^2\rangle}{2}d^3 r.
\end{equation}
This is equivalent to the fact that the cycle-integrated momentum and position energies of a simple harmonic oscillator must be equal.
The spatial field distribution of a mode, as governed by the Helmholtz equation and a particular set of boundary conditions, does not change when dispersion is added, so long as $\epsilon$ and $\mu$ are not changed for the modal frequency.  Thus \eq{Eq:yin=yang} remains true when dispersion is taken into account (although it no longer equates the total electric and magnetic contributions to the modal energy).

Applying \eq{Eq:yin=yang} to an integral of \eq{Eq:u-lossless} over the modal volume of a homogeneous cavity gives
\begin{equation}\label{Eq:U2}
 U=\left(1+\frac{1}{2}\frac{d \ln \epsilon}{d\ln \omega}+\frac{1}{2}\frac{d \ln \mu}{d \ln \omega}\right)
          \int u_f d^3 r,
\end{equation} 
where $U$ is the total modal energy and $u_f$ is the field energy density given by \eq{Eq:uf}.
Because we are taking $\mu$ and $\epsilon$ and their first derivatives to be real, we can substitute \eq{Eq:ng-over-n} into \eq{Eq:U2}, which gives
\begin{equation}\label{Eq:U3}
U=\frac{n_g}{n}U_f,
\end{equation} 
where the field energy $U_f$ is defined by
\begin{equation}\label{Eq:Uf}
 U_f=\int\left(\frac{\epsilon\langle E^2\rangle}{2}+\frac{\mu\langle H^2 \rangle}{2}\right)d^3 r.
\end{equation} 
\eq{Eq:U3} is a modal energy analog to \eq{Eq:u-from-uf} for planar waves.

$U_f$ is a cycle-averaged expression for the energy of a nondispersive electromagnetic mode.  \eq{Eq:U3} gives a simple description of the effect of dispersion on the total modal energy associated with a given field strength.

\subsection{The quantized electric field}
Assuming that a dispersive cavity mode may be quantized as a simple harmonic oscillator, we write its Hamiltonian $H$ in terms of an annihilation operator ($a$) and a creation operator ($a^\dagger$) as
\begin{equation}\label{Eq:Hoscillator}
 H=\hbar \omega\left(a^\dagger a+\frac{1}{2}\right).
\end{equation}
We note here that the energy associated with $H$ cannot be negative and cannot therefore be related to \eq{Eq:U3} when the ratio $n_g/n$ is negative \footnote{Interestingly, a second argument suggests the impossibility of a lossless resonance at a frequency where there is a negative group velocity.  Causality dictates that the combination of losslessness and a negative group velocity is only seen in an active medium.  If there is a frequency with a negative group velocity, then nearby are at least two frequencies that share the same wavelength as the on-resonance frequency and that are therefore also at resonance.  However, it appears that at least one of these must also be at a frequency where there is gain.  Net round trip gain at resonance in a cavity is not tenable in the steady state: the cavity must either lase or have its quality destroyed.}.
The electric field may then be written in terms of $a$ and $a^\dagger$ in the form
\begin{equation}\label{Eq:E}
 \bE(\br, t)=\mathcal{E} a {\bf f}(\br) e^{i\omega t}+\mathcal{E}^* a^\dagger {\bf f}^*(\br) e^{-i\omega t},
\end{equation} 
where ${\bf f}(\br)$ gives the spatial distribution and polarization of the electric field and $\mathcal{E}$ is a complex number whose amplitude gives a characteristic electric field strength.  
To make the effect of the mode volume ($V$) on the electric field explicit, we choose to normalize ${\bf f}$ according to the rule $\int{f^2 d^3 r}=V$. Combining
Eqs. ~(\ref{Eq:E}), ~(\ref{Eq:Hoscillator}), ~(\ref{Eq:Uf}), ~(\ref{Eq:U3}), and ~(\ref{Eq:yin=yang}) then gives
\begin{equation}\label{Eq:FS}
 |\mathcal{E}|=\sqrt{\frac{\hbar \omega n}{\epsilon n_g V}}.
\end{equation} 
If we restrict our consideration to dielectrics, then $\epsilon=n^2$, and $|\mathcal{E}|=\sqrt{\frac{\hbar \omega}{\epsilon_0 n n_g V}}$, which is the expression for the electric field strength given by Garrison and Chiao \cite{Garrison-Chiao-2008} who follow Milonni \cite{Milonni-1995}.  This expression agrees with a more general and earlier one implicit in work by Drummond \cite{Drummond-1990}.
In addition, dispersive expressions for characteristic field strengths (electrical or otherwise) may be related to expressions that neglect dispersion \cite{Glauber-Lewenstein-1990} using the rule $|\mathcal{E}_{dispersive}|/|\mathcal{E}_{nondispersive}|=\sqrt{n/n_g}$.

\eq{Eq:FS} suggests three separate ways to increase the characteristic electric field strength: the mode volume $V$ may be decreased, the impedance $n/\epsilon$ may be increased, or the group index $n_g$ may be brought close to zero.  In the ideal white light cavity, $n_g\rightarrow 0$, which implies that $|\mathcal{E}|\rightarrow \infty$.

One other interesting aspect of the ideal white light cavity is the mode energy.  As $n_g\rightarrow 0$ from the positive side, $U$ as defined in \eq{Eq:U3} also approaches zero.  This is a result of a cancellation of energy between a negative stored energy ($U_s$) and a positive field energy ($U_f$).  By analogy with the definition of $U_f$, we can define a field Hamiltonian ($H_f$) as
\begin{equation}
 H_f=\hbar \omega\frac{n}{n_g}\left(a^\dagger a+\frac{1}{2}\right),
\end{equation} 
which gives a vacuum field noise energy of
\begin{equation}
 U_v=\frac{\hbar \omega n}{2 n_g},
\end{equation} 
which becomes infinite for the closed cavity model as $n_g$ approaches zero.
In atomic vapor systems currently used to achieve anomalous dispersion, $n$, $\epsilon_r$, and $\mu_r$ are very close to $1$.  The total vacuum field noise then reduces to
\begin{equation}\label{Eq:vapor-field-energy}
 U_v=\frac{\hbar \omega}{2 n_g}.
\end{equation} 

In the next section we will present an open cavity model of a white light cavity that will reveal the effect of the group index on the cavity noise spectrum and allow us to make specific calculations based on reasonable models for a white light cavity medium.

\section{Coupled cavity approach}\label{se:open}
\begin{figure}
	\centerline{\includegraphics[width=8.5cm]{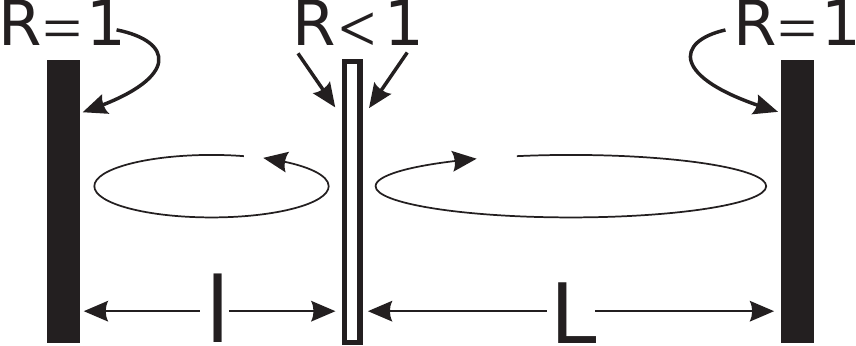}}
	\caption{\label{Fig:cavities} A small cavity of length $\ell$ is coupled to a larger cavity of length $L$ by a mirror with reflectance $R$.  The small cavity contains a lossless, dispersive medium with a group index of $n_g$.}
\end{figure}

In the previous section, we saw that the standard expression for the electromagnetic energy density in a dispersive medium implies that dispersion may scale the amplitudes of the electromagnetic fields associated with frequency eigenmodes in a closed, lossless cavity.  Because the closed cavity treatment neglects loss, the field modes in the model are delta functions in frequency and give no insight into the effect of dispersion on the power spectrum of the modes.  For the case of a white light cavity, the closed cavity treatment also suffers from the fact that the classical expression for the electromagnetic energy density of a white light mode suggests a total energy of zero.
In this section we extend and corroborate the closed cavity approach by examining the effect of dispersive on the electromagnetic fields of a lossy cavity.

Cavity loss means that the electromagnetic energy inside a cavity is coupled to some external system or systems.  
In practice, cavity modes are coupled both to external electromagnetic field modes and also to atomic systems which in turn may be coupled to other modes and systems.  Electromagnetic field modes are equivalent to simple harmonic oscillators.
In linear regimes, atomic systems may also be modeled as simple harmonic oscillators.   Thus a simple and reasonable way to model loss is to couple the system explicitly to a set of simple harmonic oscillators.  

The simplest way to do this may be to couple the lossy cavity to a larger cavity.  The lossy cavity mode then behaves as a coupled harmonic oscillator and the mode is referred to as a pseudomode.  This lossy cavity pseudomode can be probed using the modes of the overall system, which behave as uncoupled harmonic oscillators, and are often called ``true modes'' or ``universe modes \cite{Lang-Scully-Lamb-1973}.''  A nice discussion of the interrelations between pseudomodes, quasimodes, and true modes may be found in a thorough paper by Dalton and colleagues \cite{Dalton-Barnett-Garraway-2001}.  Our contribution here is to show not only that the true mode approach yields insight on the pseudomodes of a dispersive cavity, but that it does so in a way that not depend on the exact form of the electromagnetic energy density.

Figure ~\ref{Fig:cavities} depicts a simple 1 dimensional implementation of this approach involving three mirrors.  The two outer mirrors are perfect, and the combined system is closed.  
The inner mirror has a reflectance $R$, is lossless, and is frequency independent for the range of frequencies that will interest us. These three mirrors form three distinct resonant systems: a longer reservoir cavity of length $L$, a shorter cavity of length $\ell$, and a closed system comprised of the these two cavities coupled together.  From here on we will refer to these resonant systems as the \lc, the \Lc, and the \llc.  
The \Lc~is evacuated while the \lc~contains a dispersive but lossless medium with refractive index, relative permeability and relative permittivity all equal to 1 at line center.  \footnote{We choose $n=1$ not only because it is close to the refractive index value associated with many possible white light cavity implementations, but also because it allows us to make a simple distinction between total and field electromagnetic energies.}
Our object is to see how dispersion affects the pseudomodes of the \lc~by using the modes of the \llc~ as probes.

We will begin by using this model to corroborate the results of Section ~\ref{se:closed}. By neglecting absorption/gain and group velocity dispersion, we will use the open cavity model rederive \eq{Eq:vapor-field-energy} without relying on any expression for the electromagnetic energy density in a dispersive medium.  We will then extend this result by including group velocity dispersion using a specific model for an anomalously dispersive medium.

\subsection{Total field noise of an open dispersive cavity mode in the absence of loss, gain, and group velocity dispersion}
We we begin by using the model depicted in Figure ~\ref{Fig:cavities} to calculate the total field energy in a dispersive cavity pseudo-mode.  The total field energy ($U_{f\ell}$) of an \lc~pseudo-mode is a combination of contributions from many \llc~modes.  Thus
\begin{equation}
 U_{f\ell}=\ell A \sum_{\phi_\ell=-\pi}^{\phi_\ell=\pi}u_{f \ell i},
\end{equation}
where $A$ is the mode area, $u_{f\ell i}$ is the average energy density of the $i$'th \llc~mode in the \lc, and the sum is taken across one \lc~free spectral range.

For the $i$'th \llc~mode, the electric field amplitude in the \lc~ ($\bE_{\ell i}$) is related to the electric field in the \Lc~($\bE_{L i}$) by
\begin{equation}\label{Eq:field-ratios}
 \bE_{\ell i}=\bE_{L i}\frac{t}{1-\sqrt{R}e^{-i\phi_{\ell i}}},
\end{equation} 
where $t$ is the transmission coefficient of the central mirror and $\phi_{\ell i}$ is the round trip phase of the \lc~at $\omega_i$.

Using \eq{Eq:field-ratios} and noting that the ratio of \emph{field} energy densities is proportional to the squared modulus of the squared electromagnetic fields, we obtain
\begin{equation}\label{Eq:density-ratio}
 \frac{u_{f \ell i}}{u_{L i}}=\frac{1-R}{1+R-2\sqrt{R}\cos(\phi_i)},
\end{equation} 
where we use the losslessness of the mirror to get $tt*=1-R$.

\begin{equation}
 U_{f\ell}=\ell A \sum_{\phi=-\pi}^{\phi=\pi}u_{L i}\frac{1-R}{1+R-2\sqrt{R}\cos(\phi_i)}.
\end{equation} 
We know that if $R$ were $1$, the modes of the \Lc~would act as ordinary simple harmonic oscillators because we understand the quantization of electromagnetic fields in a vacuum.  The \llc~modes are perturbed by the effect of the small cavity.  However, assuming that $L$ is large and that this perturbation would therefore not change the simple harmonic oscillator character of the overall modes, we take the total vacuum energy of the overall cavity modes to be $\hbar \omega/2$.

To get an expression for $u_{L i}$, we divide by a weighted volume.  If we take the mode area in each cavity to be $A$ then the volumes of the \Lc ~and the \lc ~are $AL$ and $A\ell$.  The squared field in the \lc~ differs from the that in the~\Lc by the factor given in \eq{Eq:density-ratio}.  The energy density in the small cavity differs from that of the field energy by the factor $n_g$.  Taken together, these considerations give an \Lc~ energy density of 
\begin{equation}\label{Eq:uL}
 u_{Li}=\frac{\hbar \omega_i}{2}\frac{1}{LA+\ell^\prime_i A},
\end{equation} 
where
\begin{equation}\label{Eq:ellprime}
 \ell^\prime_i\equiv \ell n_g \frac{1-R}{1+R-2\sqrt{R}\cos(\phi_i)}.
\end{equation} 

Thus, the total field energy of the pseudo-mode may be written as
\begin{equation}
 U_{f\ell}=\ell A \sum_{\phi=-\pi}^{\phi=\pi}\frac{\hbar \omega_i}{2}\frac{1}{LA+\ell^\prime A}\frac{1-R}{1+R-2\sqrt{R}\cos(\phi_i)}.
\end{equation}

As $L$ grows large, the free spectral range of the \llc~($\Delta \omega_{L+\ell}$) falls below the minimum relevant resolution of the system and we may replace the sum over modes by a spectral integral without sacrificing accuracy.  Thus,
\begin{equation}\label{Eq:U4}
\begin{split}
 U_{f\ell}\approx \ell A \int_{\phi=-\pi}^{\phi=\pi}
                               &\frac{\hbar \omega}{2}\frac{1}{LA+\ell^\prime(\omega) A} \times \\
                  &\frac{1-R}{1+R-2\sqrt{R}\cos(\phi(\omega))}\frac{1}{\Delta\omega_{L+\ell}}d\omega.
\end{split}
\end{equation}

We can calculate $\Delta \omega_{L+\ell}$ by noting that the round-trip phase $\phi$ for any cavity must change by $2\pi$ between two adjacent resonances, giving
\begin{equation}\label{Eq:FSRexp}
 \Delta \phi=2\pi=\phi_{n+1}-\phi_{n}=\Delta \omega\frac{d\phi}{d\omega}+\frac{\Delta \omega^2}{2}\frac{d^2 \phi}{d\omega^2}+ \dots~.
\end{equation}
Since $\Delta \omega_{L+\ell}$ is small for large $L$, \eq{Eq:FSRexp} reduces to 
\begin{equation}\label{Eq:FSRapprox}
\Delta \omega_{L+\ell} \approx 2 \pi d\omega_{L+\ell}/d\phi_{L+\ell}
\end{equation} 
 when applied to the \llc.
Using the definition of the group index and the \llc~geometry, we find
\begin{equation}\label{Eq:dphi}
 \frac{d\phi_{L+\ell}}{d\omega_{L+\ell}}=\frac{2 L}{c}+\frac{2 \ell}{c} n_g\frac{1-R}{1+R-2\sqrt{R}\cos(\phi_\ell)}.
\end{equation} 
Combining Eqs. ~(\ref{Eq:FSRapprox}) and ~(\ref{Eq:dphi}) and using the definition \ref{Eq:ellprime} gives
\begin{equation}\label{Eq:FSRfinal}
 \Delta \omega_{L+\ell} \approx \frac{2 \pi c}{2(L+\ell^\prime(\omega))}.
\end{equation}

Substituting \eq{Eq:FSRfinal} into \eq{Eq:U4} gives
\begin{equation}\label{Eq:U5}
  U_{f\ell}\approx 
    \frac{\ell}{\pi c} 
    \int_{\phi=-\pi}^{\phi=\pi}
      \frac{\hbar \omega}{2}
      \frac{1-R}{1+R-2\sqrt{R}\cos(\phi(\omega))}d\omega.
\end{equation} 
We notice here that the term $L+\ell^\prime(\omega)$ resulting from the expression for $d\phi_c/d\omega_c$ cancels exactly with a similar effective length in the expression for the energy density.  This is a particular example of what may be a more general correspondence between the derivative $d\phi/d\omega$ and the electromagnetic storage capacity of lossless dispersive elements.

To simplify \eq{Eq:U5} we now make the transformation
\begin{equation}
 d\omega\rightarrow\frac{c}{2 \ell n_g} d\phi
\end{equation} 
to obtain
\begin{equation}\label{Eq:U6}
U_{f\ell}\approx 
          \frac{1}{2 \pi}
          \int_{\phi=-\pi}^{\phi=\pi}
          \frac{\hbar \omega}{2 n_g}
          \frac{1-R}{1+R-2\sqrt{R}\cos(\phi(\omega))}
          d\phi.
\end{equation} 

To further simplify \eq{Eq:U6}) we make a third approximation by pulling $\omega$ and $n_g$ out of the integral.
 Doing so gives us
\begin{equation}
U_{f\ell}\approx 
      \frac{\hbar \omega_0}{2n_g} 
          \frac{1}{2\pi}
          \int_{\phi=-\pi}^{\phi=\pi}
          \frac{1-R}{1+R-2\sqrt{R}\cos(\phi)} 
          d\phi,
\end{equation} 
and since
\begin{equation}
 \int_{\phi=-\pi}^{\phi=\pi}
          \frac{1-R}{1+R-2\sqrt{R}\cos(\phi)} 
          d\phi=2\pi,
\end{equation} 
we find that
\begin{equation}\label{Eq:noise}
U_{f\ell}\rightarrow \frac{\hbar \omega}{2 n_g},                                               
\end{equation}
which is just a restatement of \eq{Eq:vapor-field-energy}.

The preceding derivation leading to \eq{Eq:noise} may be altered as follows to avoid reliance on any particular expression for the total electromagnetic energy density in a dispersive medium.
Equation~(\ref{Eq:uL}) gives the energy density of a \llc~mode in the \Lc assuming a particular form for the energy density in the \lc.  Rewriting \eq{Eq:uL} without specifying a dispersive energy density gives
\begin{equation}\label{Eq:uL2}
 u_{Li}=\frac{\hbar \omega_i}{2}\frac{1}{LA+\ell A \eta},
\end{equation} 
where we make no assumptions on $\eta$, other than that it is finite.  As $L$ grows large, any finite value for $\eta$ eventually becomes irrelevant:
\begin{equation}\label{Eq:uLi_bigL}
 \lim_{L\to\infty}u_{Li}=\frac{\hbar \omega_i}{2}\frac{1}{LA}.
\end{equation} 
Through an identical process, the free spectral range of the \llc, given by \eq{Eq:FSRapprox}, becomes 
\begin{equation}\label{Eq:Delta_omega_bigL}
 \lim_{L\to\infty}\Delta \omega_{L+\ell} = \frac{2 \pi c}{2L}.
\end{equation} 
Limits ~(\ref{Eq:uLi_bigL}) and ~(\ref{Eq:Delta_omega_bigL}) may be used in place of Eqs. ~(\ref{Eq:uL}) and ~(\ref{Eq:FSRfinal}) with no change to the overall argument, yielding the final result given by \eq{Eq:noise} without invoking a particular form the energy density of an anomalously dispersive medium.

\eq{Eq:noise} is valid only where loss/gain and group velocity dispersion are both negligible over the cavity resonance.  For media with substantial dispersion, this limits its applicability to narrow spectral ranges.  For cavities exhibiting strong normal dispersion, this presents little difficulty because dispersion may narrow the resonance bandwidth of an already high-Q cavity.  
However, for the white light cavity in particular, $n_g\rightarrow 0$ and the vacuum field energy predicted in \eq{Eq:noise} diverges.  The open cavity model shows that this divergence comes is not due to higher buildup, but to a diverging bandwidth.  In practice, the bandwidth of a given white light resonance is limited by group velocity dispersion (GVD) and by the cavity finesse.  In the next subsection we explore the relationship between these two quantities in a specific white-light cavity realization.

\subsection{Cavity noise in a symmetric gain-doublet system}
Causality dictates that every anomalously dispersive cavity will exhibit both group velocity dispersion (GVD) and loss and/or gain.  These quantities could vanish at line center and so might be ignored if the resonance linewidth were sufficiently narrow.  As an anomalously dispersive cavity more closely approaches a white light cavity, the cavity linewidth grows.  A white light cavity would correspond to an infinitely wide linewidth.  Any physically reasonable model of a white light cavity resonance must include GVD.

A proper consideration of the total cavity noise must also account for the gain of the white light medium.  Causality dictates that a medium that is both anomalously dispersive and transparent at a frequency $\omega_o$ must exhibit gain somewhere not too distant from $\omega_o$.  This gain may be difficult to maintain in a high-finesse cavity without inducing lasing.  The difficulty may not be circumvented by making mirror reflectivity spectrally dependent without altering the anomalously dispersive nature of the cavity mode.  This gain adds noise to our system.  However, in the following calculations we will ignore its effect to show that anomalous dispersion alone is sufficient to require significant noise increases in a white light cavity.  Because we will ignore gain in the paragraphs to come, the analysis contained in them will underestimate the total noise associated with the cavity mode.

\begin{figure}
 \centerline{\includegraphics[width=9cm]{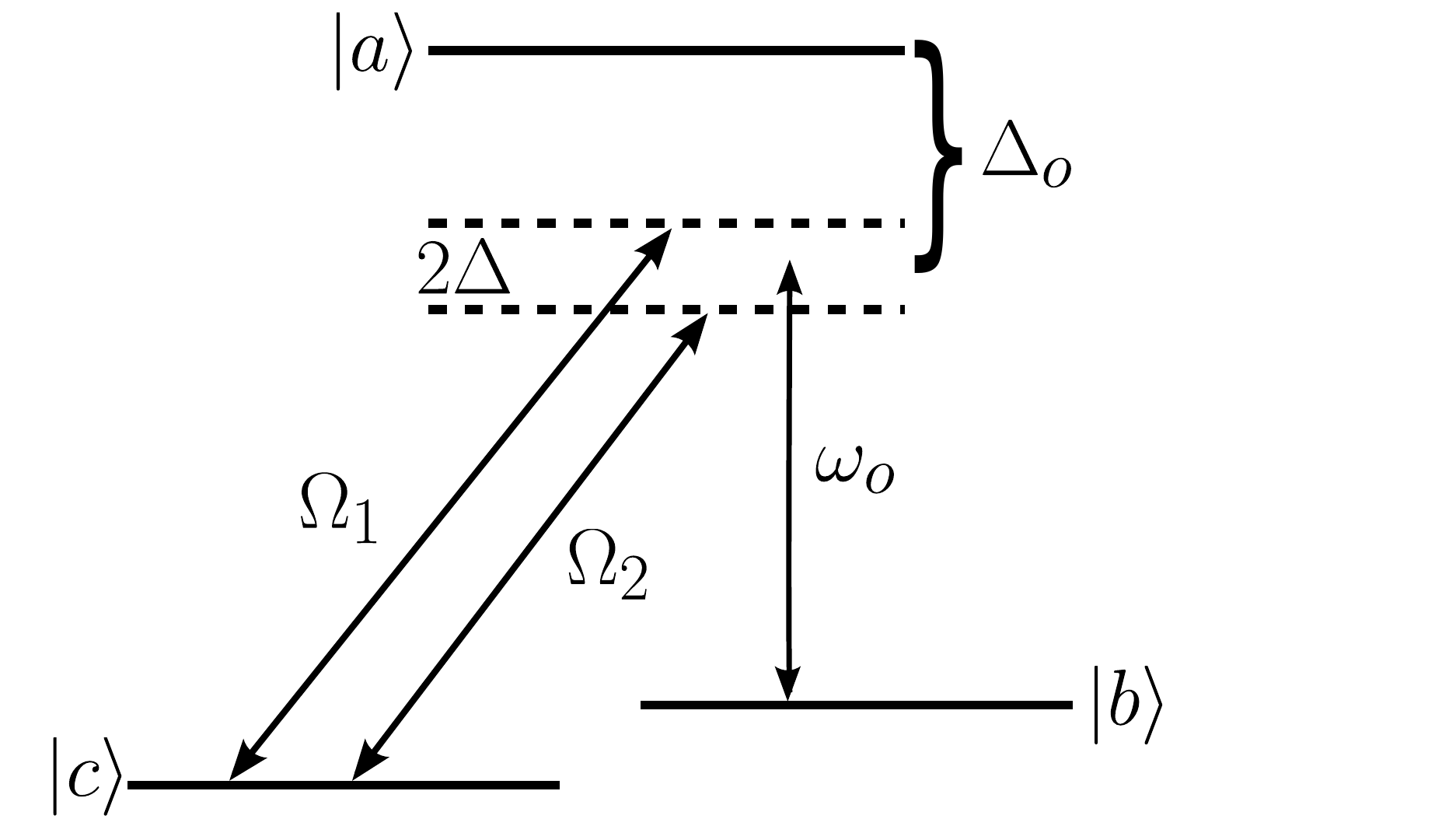}}
 \caption{\label{Fig:chi} Energy level diagram for the bichromatic Raman system.  Pump lasers represented by lines $\Omega_1$ and $\Omega_2$ provide symmetric gain lines around $\omega_o$.  Detuning from $|a\rangle$ is large ($\Delta_o >>2\Delta$).}
\end{figure}

\begin{figure}
 \centerline{\includegraphics[width=9cm]{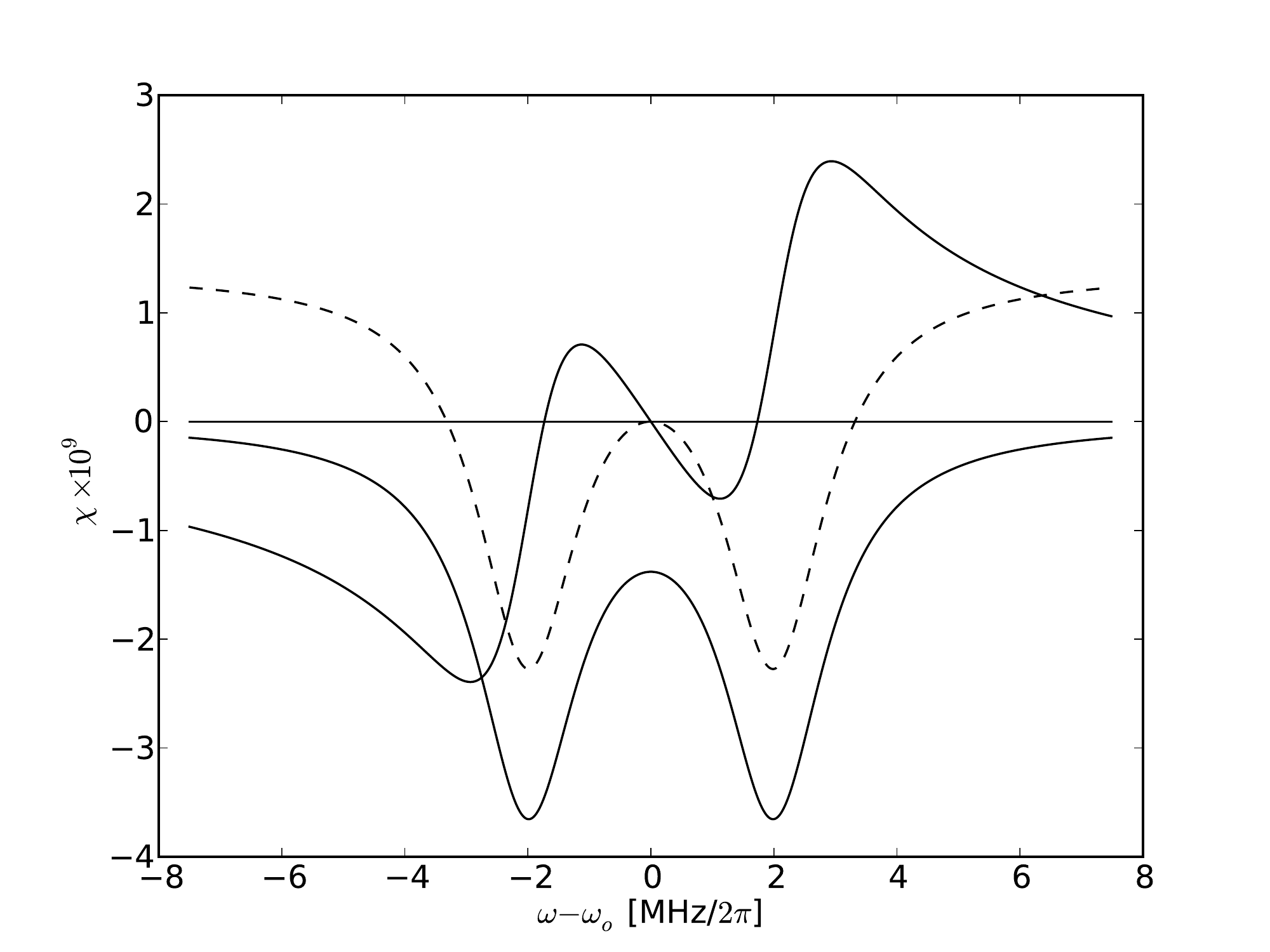}}
 \caption{\label{Fig:Raman} The susceptibility of a typical bichromatic Raman system.  The asymmetric line represents real susceptibility and the solid symmetric line represents the gain of the Raman system.  The dashed line represents the total gain in the presence of a broadband absorber whose strength is chosen to match the Raman gain at line center.}
\end{figure}

We will now solve for the pseudomode field energy given by \eq{Eq:U4} for a cavity that is homogeneously filled with a medium whose energy level diagram is depicted in Fig.~\ref{Fig:Raman}.  This medium produces a gain-doublet, as originally proposed by Steinberg and Chiao \cite{Steinberg-Chiao-1994}, accomplished through a bichromatic Raman system as experimentally realized by Wang, Kuzmich, and Dogariu \cite{Wang-Kuzmich-Dogariu-2000}.  We take the two Raman gain lines to be centered about the frequency $\omega_o$ and separated by a distance of $2\Delta$.  The value for $\Delta$ may be tuned dynamically according to experimental expedience.  The linear susceptibility ($\chi(\omega)$) may then be represented as \cite{Dogariu-Kuzmich-Wang-2001, Sun-Shahriar-Zubairy-2009} 
\begin{equation}
 \chi(\omega)=\frac{M_1}{\omega-\omega_o-\Delta+i\Gamma}+\frac{M_2}{\omega-\omega_o+\Delta+i\Gamma},
\end{equation} 
where $\Gamma$ is the Raman transition line width, and the rates $M_1$ and $M_2$ are given by
$M_j=N(|\mu_{ab}\cdot\hat e|^2/2\hbar \epsilon_o)(|\Omega_j|^2/\Delta_o^2),(j=1,2)$  \cite{Sun-Shahriar-Zubairy-2009}.
These rates $M_1$ and $M_2$ depend on $N$, the number density of participating atoms, on $\mu_{ab}\cdot \hat e$, the dipole interaction between the exciting fields and the primary Raman transition, and on the Rabi frequencies of the two exciting fields.  This last dependence means that they may be dynamically controlled by choosing the intensity of the pump lasers.  Symmetry minimizes gain and group velocity dispersion at line center when the values of $M_1$ and $M_2$ are chosen such that $M_1=M_2=M$.  For a given set of values $\omega_o$, $\Delta$, and $\Gamma$, $M$ may be chosen such that $n_g=0$ at $\omega_o$.

Figure~\ref{Fig:chi} plots the susceptibility as a function of frequency for a typical bichromatic Raman system where $M_1=M_2=M$ is chosen to minimize $n_g$ at line center. 
The dashed line indicates that this system may be combined with a broad-band absorber to roughly eliminate the net gain at line center.  
This effect was approximated in the original experiment by Wang, Kuzmich, and Dogariu \cite{Wang-Kuzmich-Dogariu-2000}.  
Loss similarly played an important role in the more recent experiments reported by Pati \emph{et al.} \cite{Pati-Salit-Salit-Shahriar-2007}.

\begin{figure}
 \centerline{\includegraphics[width=9cm]{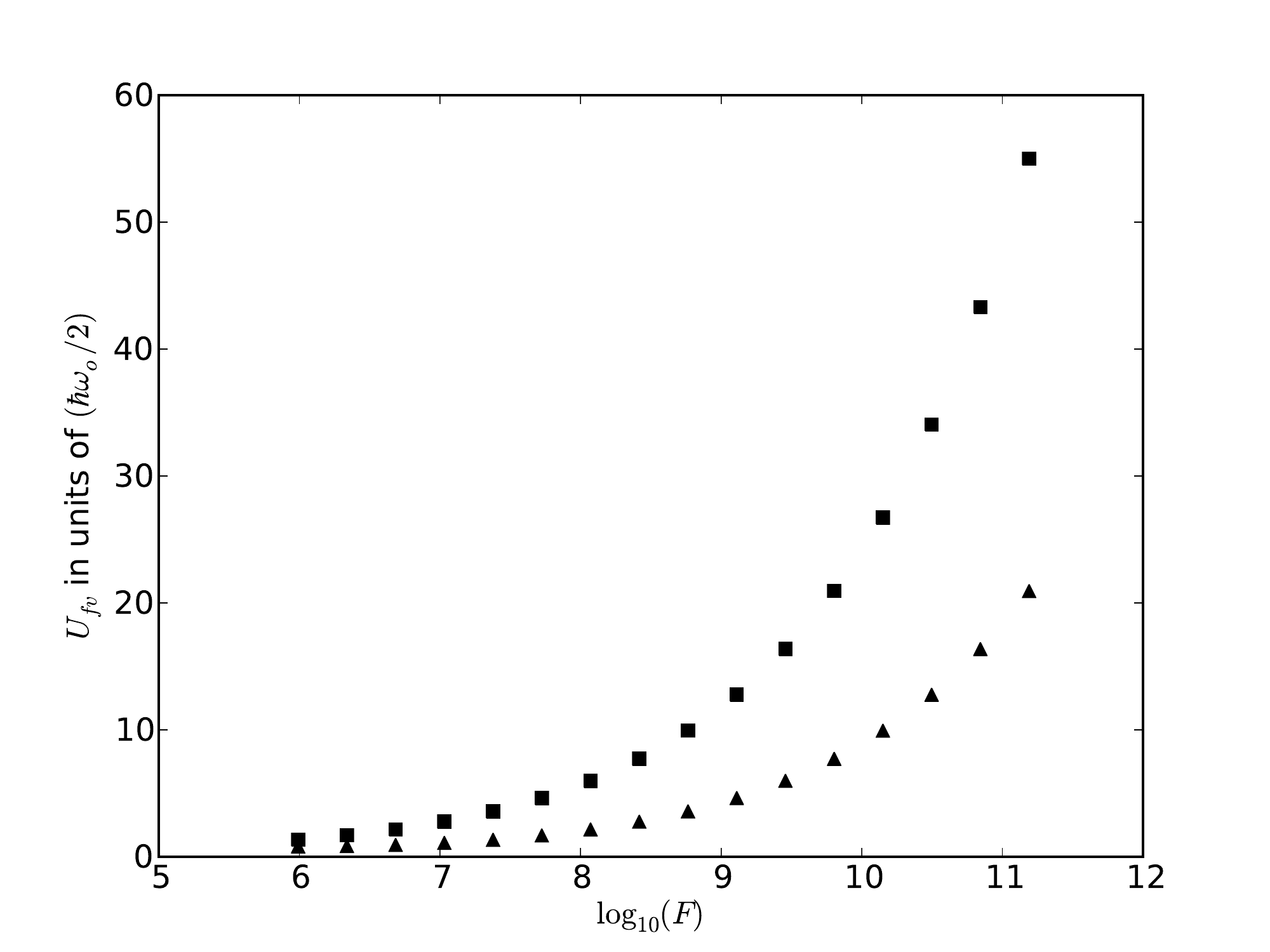}}
 \caption{\label{Fig:finesse} $U_{fv}$ is calculated for cavity lengths of 1m (squares) and 0.25m (triangles) for varying values of finesse using \eq{Eq:U5}}
\end{figure}

Figure~\ref{Fig:finesse} shows total cavity vacuum field noise for increasing finesse values as calculated from \eq{Eq:U5}.  
For both series of points, we used the values: $\Gamma=1\times 10^6$Hz/$2\pi$, $\Delta=2\times 10^6$Hz/$2\pi$, $M\approx 3.45\times 10^{-3}$Hz/$2\pi$, $\omega_o\approx 2.42\times 10^15$Hz/$2\pi$, which were taken from the literature \cite{Pati-Salit-Salit-Shahriar-2007,Sun-Shahriar-Zubairy-2009} and then adapted to avoid negative group velocities near the gain lines (Negative group velocities near the gain lines lead to extra resonances in the wings.  Causality dictates that these resonances exist at frequencies where the gain is higher or the loss lower than at line center and leads to unnecessarily large noise terms.  In many cases, it is also expected to lead to lasing.  This effect rules out the steady state stability of a large class of anomalously dispersive cavities.) We see that for both cavities the noise increases as higher finesse confines resonance to frequencies where group velocity dispersion has less effect.  As finesse becomes infinite, so does $U_{fv}$ despite the fact that \eq{Eq:U5} neglects any contribution of the gain of the active medium to the survival factor.

\subsection{Anomalous dispersion, spontaneous emission, and laser linewidth}
For a dipole emitter coupled to a large number of electromagnetic modes, the rate of spontaneous emission into one particular mode, for example the cavity mode, is proportional to the square of the electric field associated with that cavity mode.  Then \eq{Eq:E} suggests that the spontaneous emission rate into a particular cavity mode will scale with the squared electric field, which scales inversely with the group index.  When a narrow-band approximation applies, this is the case.  Thus, \eq{Eq:noise} provides a simple explanation for line narrowing Agarwal predicted in the case of lasing without inversion \cite{Agarwal-1991}; the line width is already sufficiently narrow compared to the linewidth of the emitting particle such that further narrowing provides a simple effect on the spontaneous emission rate and the laser linewidth will scale with the group velocity.  This result, although not explicitly applied in Agarwal's paper, can explain his results. However, the result is not new. For example, the role of the group velocity in determining laser linewidth was understood by Henry \cite{Henry-1982}, who framed the laser linewidth in terms of a ratio between the spontaneous emission rate and the number of photons in a mode and got an expression for the laser linewidth which is implicitly proportional to the group velocity.  

This same result applies to anomalously dispersive resonances, provided that the cavity linewidth is still narrow compared to the line shape of the gain medium.  If this condition is not met, then the broadening due to anomalous dispersion is limited by the line shape of the gain medium.  In any case, anomalous dispersion leads to an increase in laser linewidth just as it does in the linewidth of an empty cavity.

The difference between our prediction for the effect of anomalous dispersion on laser linewidth and that of Shahriar \emph{et al.} can be remedied if we redo their calculation while taking into account the effect of anomalous dispersion on the cavity photon lifetime.  Following previous authors\cite{Dorschner-Haus-Statz-1980}, they used a formula for the quantum limited linewidth that is equivalent to
\begin{equation}
 \Delta \omega_{laser}=\frac{1}{\tau_c}\sqrt{n},
\end{equation}
where $\tau_c$ was a photon lifetime and $n$.  They took $\tau_c$ to be the photon lifetime of the evacuated cavity.  However, the photon lifetime for the anomalously dispersive cavity differs from that of the evacuated cavity in their case by the factor $1/n_g$, where $n_g$ is a cavity averaged group index.  If this extra factor is taken into account, then their prediction agrees with ours.

\section{Conclusion}
We have shown that the benefits achieved through intracavity anomalous dispersion, i.e. high bandwidth combined with high buildup, come only at the cost of increased electromagnetic field noise.  An infinite finesse combined with a perfect white light mode would lead to an infinite amount of field energy in the ground state of the cavity mode.  Relaxing the infinite finesse condition to allow for cavity loss broadens the mode and relaxes the divergence, but still suggests a substantial increase in quantum field noise.

We would like to thank Ivan Deutsch and Peter Milonni and Karen Tate for help in the development of this work.  In addition, we express gratitude for a Los Alamos Laboratory Directed Research and Development grant, which funded much of this work.


\begin{thebibliography}{27}
\expandafter\ifx\csname natexlab\endcsname\relax\def\natexlab#1{#1}\fi
\expandafter\ifx\csname bibnamefont\endcsname\relax
  \def\bibnamefont#1{#1}\fi
\expandafter\ifx\csname bibfnamefont\endcsname\relax
  \def\bibfnamefont#1{#1}\fi
\expandafter\ifx\csname citenamefont\endcsname\relax
  \def\citenamefont#1{#1}\fi
\expandafter\ifx\csname url\endcsname\relax
  \def\url#1{\texttt{#1}}\fi
\expandafter\ifx\csname urlprefix\endcsname\relax\def\urlprefix{URL }\fi
\providecommand{\bibinfo}[2]{#2}
\providecommand{\eprint}[2][]{\url{#2}}

\bibitem[{\citenamefont{Englund et~al.}(2005)\citenamefont{Englund, Fattal,
  Waks, Solomon, Zhang, Nakaoka, Arakawa, Yamamoto, and
  Vu\v{c}kovi\'{c}}}]{Englund-Fattal-Vuckovic-2005}
\bibinfo{author}{\bibfnamefont{D.}~\bibnamefont{Englund}},
  \bibinfo{author}{\bibfnamefont{D.}~\bibnamefont{Fattal}},
  \bibinfo{author}{\bibfnamefont{E.}~\bibnamefont{Waks}},
  \bibinfo{author}{\bibfnamefont{G.}~\bibnamefont{Solomon}},
  \bibinfo{author}{\bibfnamefont{B.}~\bibnamefont{Zhang}},
  \bibinfo{author}{\bibfnamefont{T.}~\bibnamefont{Nakaoka}},
  \bibinfo{author}{\bibfnamefont{Y.}~\bibnamefont{Arakawa}},
  \bibinfo{author}{\bibfnamefont{Y.}~\bibnamefont{Yamamoto}}, \bibnamefont{and}
  \bibinfo{author}{\bibfnamefont{J.}~\bibnamefont{Vu\v{c}kovi\'{c}}},
  \bibinfo{journal}{Physical Review Letters} \textbf{\bibinfo{volume}{95}},
  \bibinfo{pages}{013904+} (\bibinfo{year}{2005}).

\bibitem[{\citenamefont{G\'{e}rard et~al.}()\citenamefont{G\'{e}rard,
  Gregersen, Nielsen, Claudon, and Moerk}}]{Moerk-2010}
\bibinfo{author}{\bibfnamefont{J.~M.} \bibnamefont{G\'{e}rard}},
  \bibinfo{author}{\bibfnamefont{N.}~\bibnamefont{Gregersen}},
  \bibinfo{author}{\bibfnamefont{T.~R.} \bibnamefont{Nielsen}},
  \bibinfo{author}{\bibfnamefont{J.}~\bibnamefont{Claudon}}, \bibnamefont{and}
  \bibinfo{author}{\bibfnamefont{J.}~\bibnamefont{Moerk}} (2010).

\bibitem[{\citenamefont{Candler}(1946)}]{Candler-1946}
\bibinfo{author}{\bibfnamefont{A.~C.} \bibnamefont{Candler}},
  \bibinfo{journal}{Nature} \textbf{\bibinfo{volume}{157}},
  \bibinfo{pages}{444} (\bibinfo{year}{1946}).

\bibitem[{\citenamefont{Wicht et~al.}(2000)\citenamefont{Wicht, M\"{u}ller,
  Rinkleff, Rocco, and Danzmann}}]{Wicht-Muller-Rinkleff-Danzmann-2000}
\bibinfo{author}{\bibfnamefont{A.}~\bibnamefont{Wicht}},
  \bibinfo{author}{\bibfnamefont{M.}~\bibnamefont{M\"{u}ller}},
  \bibinfo{author}{\bibfnamefont{R.~H.} \bibnamefont{Rinkleff}},
  \bibinfo{author}{\bibfnamefont{A.}~\bibnamefont{Rocco}}, \bibnamefont{and}
  \bibinfo{author}{\bibfnamefont{K.}~\bibnamefont{Danzmann}},
  \bibinfo{journal}{Optics Communications} \textbf{\bibinfo{volume}{179}},
  \bibinfo{pages}{107} (\bibinfo{year}{2000}).

\bibitem[{\citenamefont{Pati et~al.}(2007)\citenamefont{Pati, Salit, Salit, and
  Shahriar}}]{Pati-Salit-Salit-Shahriar-2007}
\bibinfo{author}{\bibfnamefont{G.~S.} \bibnamefont{Pati}},
  \bibinfo{author}{\bibfnamefont{M.}~\bibnamefont{Salit}},
  \bibinfo{author}{\bibfnamefont{K.}~\bibnamefont{Salit}}, \bibnamefont{and}
  \bibinfo{author}{\bibfnamefont{M.~S.} \bibnamefont{Shahriar}},
  \bibinfo{journal}{Physical Review Letters} \textbf{\bibinfo{volume}{99}}
  (\bibinfo{year}{2007}).

\bibitem[{\citenamefont{Wicht et~al.}(1997)\citenamefont{Wicht, Danzmann,
  Fleischhauer, Scully, M\"{u}ller, and
  Rinkleff}}]{Wicht-Danzmann-Rinkleff-1997}
\bibinfo{author}{\bibfnamefont{A.}~\bibnamefont{Wicht}},
  \bibinfo{author}{\bibfnamefont{K.}~\bibnamefont{Danzmann}},
  \bibinfo{author}{\bibfnamefont{M.}~\bibnamefont{Fleischhauer}},
  \bibinfo{author}{\bibfnamefont{M.}~\bibnamefont{Scully}},
  \bibinfo{author}{\bibfnamefont{G.}~\bibnamefont{M\"{u}ller}},
  \bibnamefont{and} \bibinfo{author}{\bibnamefont{Rinkleff}},
  \bibinfo{journal}{Optics Communications} \textbf{\bibinfo{volume}{134}},
  \bibinfo{pages}{431} (\bibinfo{year}{1997}).

\bibitem[{\citenamefont{Wise et~al.}(2004)\citenamefont{Wise, Mueller, Reitze,
  Tanner, and Whiting}}]{Wise-Whiting-2004}
\bibinfo{author}{\bibfnamefont{S.}~\bibnamefont{Wise}},
  \bibinfo{author}{\bibfnamefont{G.}~\bibnamefont{Mueller}},
  \bibinfo{author}{\bibfnamefont{D.}~\bibnamefont{Reitze}},
  \bibinfo{author}{\bibfnamefont{D.~B.} \bibnamefont{Tanner}},
  \bibnamefont{and} \bibinfo{author}{\bibfnamefont{B.~F.}
  \bibnamefont{Whiting}}, \bibinfo{journal}{Classical and Quantum Gravity}
  \textbf{\bibinfo{volume}{21}}, \bibinfo{pages}{S1031} (\bibinfo{year}{2004}).

\bibitem[{\citenamefont{Karapetyan}(2004)}]{Karapetyan-2004}
\bibinfo{author}{\bibfnamefont{G.~G.} \bibnamefont{Karapetyan}},
  \bibinfo{journal}{Optics Communications} \textbf{\bibinfo{volume}{238}},
  \bibinfo{pages}{35} (\bibinfo{year}{2004}).

\bibitem[{\citenamefont{Shahriar et~al.}(2007)\citenamefont{Shahriar, Pati,
  Tripathi, Gopal, Messall, and
  Salit}}]{Shahriar-Pati-Tripathy-Gopal-Messall-Salit-2007}
\bibinfo{author}{\bibfnamefont{M.~S.} \bibnamefont{Shahriar}},
  \bibinfo{author}{\bibfnamefont{G.~S.} \bibnamefont{Pati}},
  \bibinfo{author}{\bibfnamefont{R.}~\bibnamefont{Tripathi}},
  \bibinfo{author}{\bibfnamefont{V.}~\bibnamefont{Gopal}},
  \bibinfo{author}{\bibfnamefont{M.}~\bibnamefont{Messall}}, \bibnamefont{and}
  \bibinfo{author}{\bibfnamefont{K.}~\bibnamefont{Salit}},
  \bibinfo{journal}{Physical Review A (Atomic, Molecular, and Optical Physics)}
  \textbf{\bibinfo{volume}{75}} (\bibinfo{year}{2007}).

\bibitem[{\citenamefont{Sun et~al.}(2009)\citenamefont{Sun, Shahriar, and
  Zubairy}}]{Sun-Shahriar-Zubairy-2009}
\bibinfo{author}{\bibfnamefont{Q.}~\bibnamefont{Sun}},
  \bibinfo{author}{\bibfnamefont{M.~S.} \bibnamefont{Shahriar}},
  \bibnamefont{and} \bibinfo{author}{\bibfnamefont{M.~S.}
  \bibnamefont{Zubairy}} (\bibinfo{year}{2009}), \eprint{0909.5443},
  \urlprefix\url{http://arxiv.org/abs/0909.5443}.

\bibitem[{\citenamefont{Wicht et~al.}(2002)\citenamefont{Wicht, Rinkleff,
  Molella, and Danzmann}}]{Wicht-Rinkleff-Danzmann-2002}
\bibinfo{author}{\bibfnamefont{A.}~\bibnamefont{Wicht}},
  \bibinfo{author}{\bibfnamefont{R.~H.} \bibnamefont{Rinkleff}},
  \bibinfo{author}{\bibfnamefont{L.~S.} \bibnamefont{Molella}},
  \bibnamefont{and} \bibinfo{author}{\bibfnamefont{K.}~\bibnamefont{Danzmann}},
  \bibinfo{journal}{Physical Review A} \textbf{\bibinfo{volume}{66}},
  \bibinfo{pages}{063815+} (\bibinfo{year}{2002}).

\bibitem[{\citenamefont{Landau et~al.}(1984)\citenamefont{Landau, Pitaevskii,
  and Lifshitz}}]{Landau-Pitaevski-Lifshitz}
\bibinfo{author}{\bibfnamefont{L.~D.} \bibnamefont{Landau}},
  \bibinfo{author}{\bibfnamefont{L.~P.} \bibnamefont{Pitaevskii}},
  \bibnamefont{and} \bibinfo{author}{\bibfnamefont{E.~M.}
  \bibnamefont{Lifshitz}}, \emph{\bibinfo{title}{Electrodynamics of Continuous
  Media, Second Edition: Volume 8 (Course of Theoretical Physics)}}
  (\bibinfo{publisher}{Butterworth-Heinemann}, \bibinfo{year}{1984}),
  \bibinfo{edition}{2nd} ed., ISBN \bibinfo{isbn}{0750626348}.

\bibitem[{\citenamefont{Jackson}(1998)}]{Jackson-1998}
\bibinfo{author}{\bibfnamefont{J.~D.} \bibnamefont{Jackson}},
  \emph{\bibinfo{title}{Classical Electrodynamics Third Edition}}
  (\bibinfo{publisher}{Wiley}, \bibinfo{year}{1998}), \bibinfo{edition}{3rd}
  ed., ISBN \bibinfo{isbn}{047130932X}.

\bibitem[{\citenamefont{Peatross et~al.}(2001)\citenamefont{Peatross, Ware, and
  Glasgow}}]{Peatross-Ware-Glasgow-2001}
\bibinfo{author}{\bibfnamefont{J.}~\bibnamefont{Peatross}},
  \bibinfo{author}{\bibfnamefont{M.}~\bibnamefont{Ware}}, \bibnamefont{and}
  \bibinfo{author}{\bibfnamefont{S.~A.} \bibnamefont{Glasgow}},
  \bibinfo{journal}{J. Opt. Soc. Am. A} \textbf{\bibinfo{volume}{18}},
  \bibinfo{pages}{1719} (\bibinfo{year}{2001}).

\bibitem[{\citenamefont{Chiao and Garrison}(2008)}]{Garrison-Chiao-2008}
\bibinfo{author}{\bibfnamefont{R.}~\bibnamefont{Chiao}} \bibnamefont{and}
  \bibinfo{author}{\bibfnamefont{J.}~\bibnamefont{Garrison}},
  \emph{\bibinfo{title}{Quantum Optics (Oxford Graduate Texts)}}
  (\bibinfo{publisher}{Oxford University Press, USA}, \bibinfo{year}{2008}),
  ISBN \bibinfo{isbn}{0198508867}.

\bibitem[{\citenamefont{Milonni}(1995)}]{Milonni-1995}
\bibinfo{author}{\bibfnamefont{P.~W.} \bibnamefont{Milonni}},
  \bibinfo{journal}{Journal of Modern Optics} \textbf{\bibinfo{volume}{42}},
  \bibinfo{pages}{1991} (\bibinfo{year}{1995}).

\bibitem[{\citenamefont{Drummond}(1990)}]{Drummond-1990}
\bibinfo{author}{\bibfnamefont{P.~D.} \bibnamefont{Drummond}},
  \bibinfo{journal}{Physical Review A} \textbf{\bibinfo{volume}{42}},
  \bibinfo{pages}{6845+} (\bibinfo{year}{1990}).

\bibitem[{\citenamefont{Glauber and
  Lewenstein}(1991)}]{Glauber-Lewenstein-1990}
\bibinfo{author}{\bibfnamefont{R.~J.} \bibnamefont{Glauber}} \bibnamefont{and}
  \bibinfo{author}{\bibfnamefont{M.}~\bibnamefont{Lewenstein}},
  \bibinfo{journal}{Physical Review A} \textbf{\bibinfo{volume}{43}},
  \bibinfo{pages}{467} (\bibinfo{year}{1991}).

\bibitem[{\citenamefont{Lang et~al.}(1973)\citenamefont{Lang, Scully, and
  Lamb}}]{Lang-Scully-Lamb-1973}
\bibinfo{author}{\bibfnamefont{R.}~\bibnamefont{Lang}},
  \bibinfo{author}{\bibfnamefont{M.~O.} \bibnamefont{Scully}},
  \bibnamefont{and} \bibinfo{author}{\bibfnamefont{W.~E.} \bibnamefont{Lamb}},
  \bibinfo{journal}{Physical Review A} \textbf{\bibinfo{volume}{7}},
  \bibinfo{pages}{1788+} (\bibinfo{year}{1973}).

\bibitem[{\citenamefont{Dalton et~al.}(2001)\citenamefont{Dalton, Barnett, and
  Garraway}}]{Dalton-Barnett-Garraway-2001}
\bibinfo{author}{\bibfnamefont{B.~J.} \bibnamefont{Dalton}},
  \bibinfo{author}{\bibfnamefont{S.~M.} \bibnamefont{Barnett}},
  \bibnamefont{and} \bibinfo{author}{\bibfnamefont{B.~M.}
  \bibnamefont{Garraway}}, \bibinfo{journal}{Physical Review A}
  \textbf{\bibinfo{volume}{64}}, \bibinfo{pages}{053813+}
  (\bibinfo{year}{2001}).

\bibitem[{\citenamefont{Steinberg and Chiao}(1994)}]{Steinberg-Chiao-1994}
\bibinfo{author}{\bibfnamefont{A.~M.} \bibnamefont{Steinberg}}
  \bibnamefont{and} \bibinfo{author}{\bibfnamefont{R.~Y.} \bibnamefont{Chiao}},
  \bibinfo{journal}{Physical Review A} \textbf{\bibinfo{volume}{49}},
  \bibinfo{pages}{2071} (\bibinfo{year}{1994}).

\bibitem[{\citenamefont{Wang et~al.}(2000)\citenamefont{Wang, Kuzmich, and
  Dogariu}}]{Wang-Kuzmich-Dogariu-2000}
\bibinfo{author}{\bibfnamefont{L.~J.} \bibnamefont{Wang}},
  \bibinfo{author}{\bibfnamefont{A.}~\bibnamefont{Kuzmich}}, \bibnamefont{and}
  \bibinfo{author}{\bibfnamefont{A.}~\bibnamefont{Dogariu}},
  \bibinfo{journal}{Nature} \textbf{\bibinfo{volume}{406}},
  \bibinfo{pages}{277} (\bibinfo{year}{2000}).

\bibitem[{\citenamefont{Dogariu et~al.}(2001)\citenamefont{Dogariu, Kuzmich,
  and Wang}}]{Dogariu-Kuzmich-Wang-2001}
\bibinfo{author}{\bibfnamefont{A.}~\bibnamefont{Dogariu}},
  \bibinfo{author}{\bibfnamefont{A.}~\bibnamefont{Kuzmich}}, \bibnamefont{and}
  \bibinfo{author}{\bibfnamefont{L.~J.} \bibnamefont{Wang}},
  \bibinfo{journal}{Physical Review A} \textbf{\bibinfo{volume}{63}},
  \bibinfo{pages}{053806+} (\bibinfo{year}{2001}).

\bibitem[{\citenamefont{Agarwal}(1991)}]{Agarwal-1991}
\bibinfo{author}{\bibfnamefont{G.~S.} \bibnamefont{Agarwal}},
  \bibinfo{journal}{Physical Review Letters} \textbf{\bibinfo{volume}{67}},
  \bibinfo{pages}{980+} (\bibinfo{year}{1991}).

\bibitem[{\citenamefont{Henry}(2003)}]{Henry-1982}
\bibinfo{author}{\bibfnamefont{C.}~\bibnamefont{Henry}},
  \bibinfo{journal}{Quantum Electronics, IEEE Journal of}
  \textbf{\bibinfo{volume}{18}}, \bibinfo{pages}{259} (\bibinfo{year}{2003}).

\bibitem[{\citenamefont{Dorschner et~al.}(1980)\citenamefont{Dorschner, Haus,
  Holz, Smith, and Statz}}]{Dorschner-Haus-Statz-1980}
\bibinfo{author}{\bibfnamefont{T.}~\bibnamefont{Dorschner}},
  \bibinfo{author}{\bibfnamefont{H.}~\bibnamefont{Haus}},
  \bibinfo{author}{\bibfnamefont{M.}~\bibnamefont{Holz}},
  \bibinfo{author}{\bibfnamefont{I.}~\bibnamefont{Smith}}, \bibnamefont{and}
  \bibinfo{author}{\bibfnamefont{H.}~\bibnamefont{Statz}},
  \bibinfo{journal}{Quantum Electronics, IEEE Journal of}
  \textbf{\bibinfo{volume}{16}}, \bibinfo{pages}{1376} (\bibinfo{year}{1980}).

\bibitem[{\citenamefont{Glasgow et~al.}(2007)\citenamefont{Glasgow, Meilstrup,
  Peatross, and Ware}}]{Glasgow-Meilstrup-Peatross-Ware-2007}
\bibinfo{author}{\bibfnamefont{S.}~\bibnamefont{Glasgow}},
  \bibinfo{author}{\bibfnamefont{M.}~\bibnamefont{Meilstrup}},
  \bibinfo{author}{\bibfnamefont{J.}~\bibnamefont{Peatross}}, \bibnamefont{and}
  \bibinfo{author}{\bibfnamefont{M.}~\bibnamefont{Ware}},
  \bibinfo{journal}{Physical Review E (Statistical, Nonlinear, and Soft Matter
  Physics)} \textbf{\bibinfo{volume}{75}} (\bibinfo{year}{2007}).

\end{thebibliography}

\end{document}